%% file: main.tex
\documentclass[journal]{IEEEtran}
\usepackage{mathptmx} % This is Times font
\usepackage{fancyhdr}
\usepackage[normalem]{ulem}
\usepackage[hyphens]{url}
\usepackage[usenames, dvipsnames]{color}
\usepackage{hyperref}
\usepackage{booktabs}
\usepackage{bbding}
\usepackage{pifont}
\usepackage{graphicx}
\usepackage{multirow}
\usepackage{amsmath}
\usepackage{setspace}
\usepackage{cite}
\usepackage{siunitx}
\usepackage{threeparttable}
\usepackage[utf8]{inputenc}
\usepackage{booktabs, caption, makecell}
\usepackage{multirow}

\newcommand{\ignore}[1]{}

\newif\ifsubmit
\submitfalse
\ifsubmit
    \newcommand{\minsuk}[1]{}
    \newcommand{\kaushik}[1]{}
    \newcommand{\gopal}[1]{}
    \newcommand{\yong}[1]{}
    \newcommand{\todo}[1]{}
    \newcommand{\tocite}[1]{}
    \newcommand{\revised}[1]{}
\else
    \definecolor{comments}{rgb}{0.3, 0.7, 1.0}
    \newcommand{\minsuk}[1]{[{\color{comments}MK: #1}]}
    \newcommand{\kaushik}[1]{[{\color{comments}KR: #1}]}
    \newcommand{\gopal}[1]{[{\color{comments}GS: #1}]}
    \newcommand{\yong}[1]{[{\color{comments}YS: #1}]}
    \newcommand{\todo}[1]{[{\color{cyan}TODO: #1}]}
    \newcommand{\tocite}[1]{[{\color{red}CITE: #1}]}
    \newcommand{\revised}[1]{{\color{red} #1}}
\fi

\begin{document}
%
% paper title
% Titles are generally capitalized except for words such as a, an, and, as,
% at, but, by, for, in, nor, of, on, or, the, to and up, which are usually
% not capitalized unless they are the first or last word of the title.
% Linebreaks \\ can be used within to get better formatting as desired.
% Do not put math or special symbols in the title.
\title{sBSNN: Stochastic-Bits Enabled Binary Spiking Neural Network with On-Chip Learning for Energy Efficient Neuromorphic Computing at the Edge}

\author{Minsuk~Koo,~\IEEEmembership{Member,~IEEE,}
        Gopalakrishnan~Srinivasan,
        Yong~Shim,~\IEEEmembership{Member,~IEEE,}
        and~Kaushik~Roy,~\IEEEmembership{Fellow,~IEEE}% <-this % stops a space
\thanks{M. Koo, G. Srinivasan, and K. Roy are with the department
of Electrical and Computer Engineering, Purdue University, West Lafayette,
IN 47907, USA. E-mail: (koom, srinivg, kaushik)@purdue.edu.}% <-this % stops a space
\thanks{Y. Shim is with Intel Corporation, Hillsboro, OR 97125, USA.}

\thanks{Manuscript received April 19, 2005; revised August 26, 2015.}}

% The paper headers
\markboth{Journal of \LaTeX\ Class Files,~Vol.~14, No.~8, August~2015}%
{Shell \MakeLowercase{\textit{et al.}}: Bare Demo of IEEEtran.cls for IEEE Journals}
% The only time the second header will appear is for the odd numbered pages
% after the title page when using the twoside option.
% 
% *** Note that you probably will NOT want to include the author's ***
% *** name in the headers of peer review papers.                   ***
% You can use \ifCLASSOPTIONpeerreview for conditional compilation here if
% you desire.

% make the title area
\maketitle
\pagestyle{plain}

% As a general rule, do not put math, special symbols or citations
% in the abstract
\input{sec/0-abstract}

% no keywords

% For peer review papers, you can put extra information on the cover
% page as needed:
% \ifCLASSOPTIONpeerreview
% \begin{center} \bfseries EDICS Category: 3-BBND \end{center}
% \fi
%
% For peerreview papers, this IEEEtran command inserts a page break and
% creates the second title. It will be ignored for other modes.
\IEEEpeerreviewmaketitle

\input{sec/1-introduction}
\input{sec/2-background}

\input{sec/3-stochastic-bit}
\input{sec/4-result}
\input{sec/5-conclusion}
\input{sec/6-acknowledgement}

% conference papers do not normally have an appendix

% use section* for acknowledgment
% \section*{Acknowledgment}
% The authors would like to thank...

% trigger a \newpage just before the given reference
% number - used to balance the columns on the last page
% adjust value as needed - may need to be readjusted if
% the document is modified later
%\IEEEtriggeratref{8}
% The "triggered" command can be changed if desired:
%\IEEEtriggercmd{\enlargethispage{-5in}}

% references section

% can use a bibliography generated by BibTeX as a .bbl file
% BibTeX documentation can be easily obtained at:
% http://mirror.ctan.org/biblio/bibtex/contrib/doc/
% The IEEEtran BibTeX style support page is at:
% http://www.michaelshell.org/tex/ieeetran/bibtex/
%\bibliographystyle{IEEEtran}
% argument is your BibTeX string definitions and bibliography database(s)
%\bibliography{IEEEabrv,../bib/paper}
%
% <OR> manually copy in the resultant .bbl file
% set second argument of \begin to the number of references
% (used to reserve space for the reference number labels box)
%\begin{thebibliography}{1}
%\bibitem{IEEEhowto:kopka}
%H.~Kopka and P.~W. Daly, \emph{A Guide to \LaTeX}, 3rd~ed.\hskip 1em plus
%  0.5em minus 0.4em\relax Harlow, England: Addison-Wesley, 1999.
%\end{thebibliography}

\bibliographystyle{IEEEtran}
\bibliography{ref.bib}

\begin{IEEEbiography}[{\includegraphics[width=1in,height=1.25in,clip,keepaspectratio]{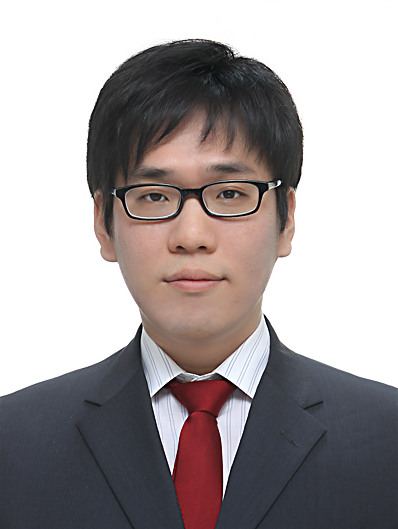}}]%
{Minsuk Koo}
received the B.S. degree in electrical
engineering from the Korea Advanced Institute of
Science and Technology (KAIST), Daejeon, Korea,
in 2007, the M.S. degree in electrical engineering
from Seoul National University, Seoul, Korea, in
2009, and is currently working toward the Ph.D.
degree in electrical and computer engineering at
Purdue University, West Lafayette, IN, USA.
From 2009 to 2012, he was with RadioPulse Inc., Seoul, Korea as a senior engineer where he had been involved with the development of ZigBee transceiver and SoC products. His research interests include circuits and system for neural networks and associative computing using CMOS and emerging devices.
 \end{IEEEbiography}

\begin{IEEEbiography}[{\includegraphics[width=1in,height=1.25in,clip,keepaspectratio]{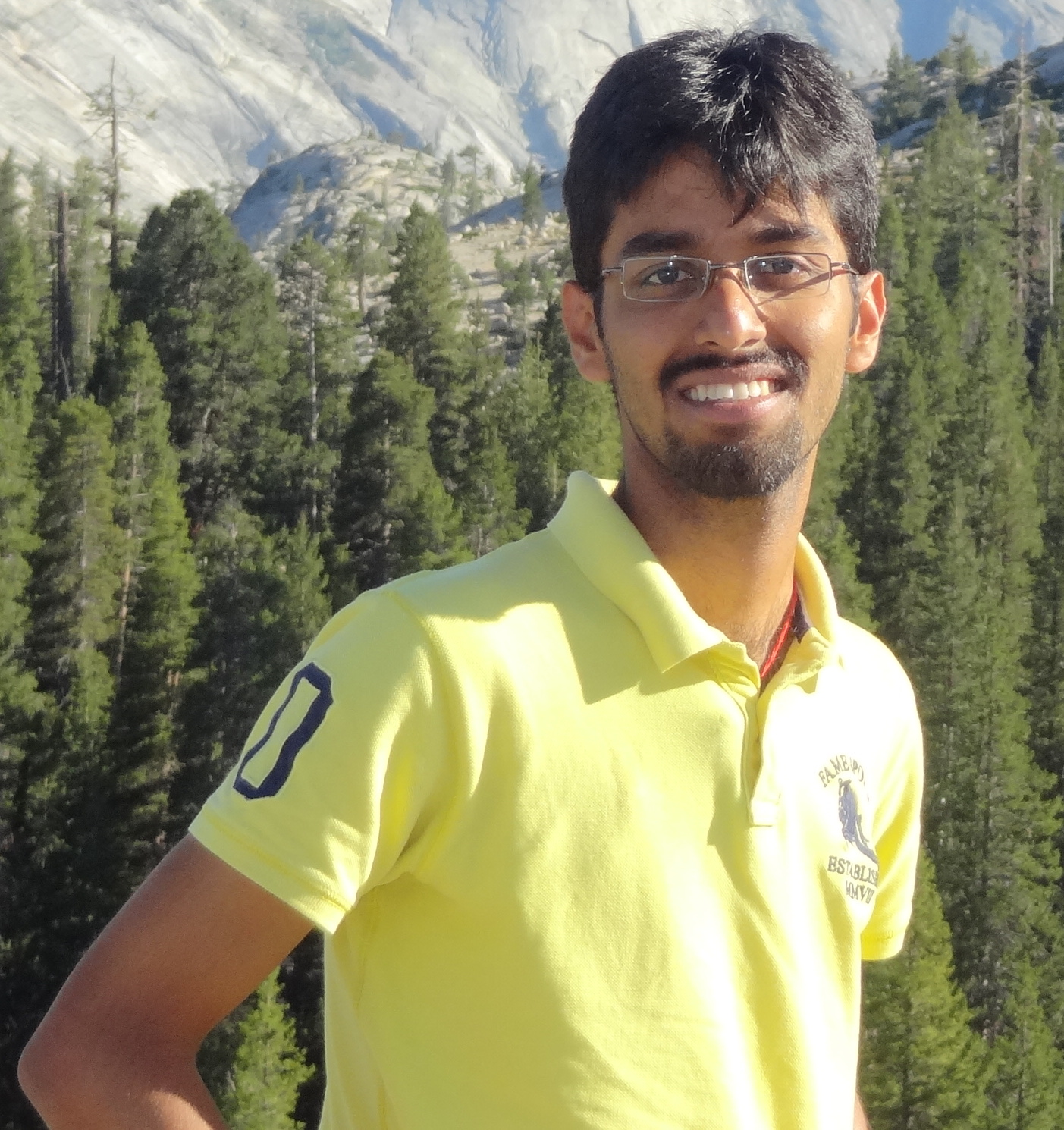}}]%
{Gopalakrishnan Srinivasan}
is currently pursuing his PhD in Electrical Engineering at Purdue University under the guidance of Prof. Kaushik Roy. His primary research interests include investigating brain inspired spiking neural network architectures and training methodologies, and their energy-efficient implementation using CMOS and post-CMOS (spintronic) technologies. He received his B.Tech. in Electrical and Electronics Engineering from the National Institute of Technology, Calicut, India, and his Masters in Computer Engineering from the North Carolina State University, Raleigh, NC, in 2010 and 2012, respectively.
\end{IEEEbiography}

\begin{IEEEbiography}[{\includegraphics[width=1in,height=1.25in,clip,keepaspectratio]{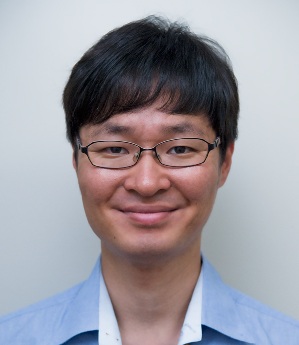}}]%
{Yong Shim}
completed his PhD at Purdue University in August 2018 and currently works as a SRAM circuit designer at Intel in Hillsboro, OR. He received the B.S. and M.S. degrees in electrical engineering from the Korea University, Seoul, Korea, in 2004 and 2006, respectively. In 2006, he joined Samsung Electronics Co., Ltd., Hwasung, Korea, where he has been involved in designing circuits for Memory Interface. He also worked as a graduate research intern at Circuit Research Labs, Intel Labs, in 2015. His research interests includes implementation of the unconventional computing models such as Neural Networks, In-memory Computing models, and optimization problem solvers based on the conventional CMOS circuits and emerging devices
\end{IEEEbiography}

\begin{IEEEbiography}[{\includegraphics[width=1in,height=1.25in,clip,keepaspectratio]{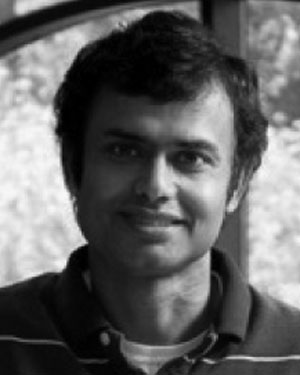}}]%
{Kaushik Roy}
(F’02) received the B.Tech. degree in
electronics and electrical communications engineering
from the Indian Institute of Technology, Kharagpur,
India, and the Ph.D. degree from the Department
of Electrical and Computer Engineering, University
of Illinois at Urbana-Champaign, Champaign,
IL, USA, in 1990. He was with the Semiconductor
Process and Design Center of Texas Instruments,
Dallas, TX, USA, where he worked on FPGA architecture
development and low-power circuit design.
He joined the electrical and computer engineering
faculty with Purdue University, West Lafayette, IN, USA, in 1993, where he is
currently Edward G. Tiedemann Jr. Distinguished Professor. He also the Director
of the Center for Brain-Inspired Computing funded by SRC/DARPA. He has
authored and coauthored more than 700 papers in refereed journals and conferences,
holds 18 patents, supervised 75 Ph.D. dissertations, and is co-author of
two books on Low Power CMOS VLSI Design (Wiley and McGraw Hill). His
research interests include neuromorphic and emerging computing models, neuromimetic
devices, spintronics, device-circuit-algorithm codesign for nanoscale
silicon and non-Silicon technologies, and low-power electronics.
Hewas the recipient of the National Science Foundation Career Development
Award in 1995, IBM faculty partnership award, ATT/Lucent Foundation award,
2005 SRC Technical Excellence Award, SRC Inventors Award, Purdue College
of Engineering Research Excellence Award, Humboldt Research Award in
2010, 2010 IEEE Circuits and Systems Society Technical Achievement Award
(Charles Doeser Award), Distinguished AlumnusAward from Indian Institute of
Technology, Kharagpur, India, Fulbright-Nehru Distinguished Chair, DoD Vannevar
Bush Faculty Fellow (2014–2019), Semiconductor Research Corporation
Aristotle award in 2015, and best paper awards at 1997 International TestConference,
IEEE 2000 International Symposium on Quality of IC Design, 2003 IEEE
Latin American Test Workshop, 2003 IEEE Nano, 2004 IEEE International
Conference on Computer Design, 2006 IEEE/ACM International Symposium
on Low Power Electronics ${\&}$ Design, and 2005 IEEE Circuits and System Society
Outstanding Young Author Award (Chris Kim), 2006 IEEE Transactions
on VLSI Systems Best Paper Award, 2012 ACM/IEEE International Symposium
on Low Power Electronics and Design Best Paper Award, 2013 IEEE
Transactions on VLSI Best Paper Award. He was a Faculty Scholar with Purdue
University (1998–2003). He was a Research Visionary Board Member of
Motorola Labs (2002) and held the M. Gandhi Distinguished Visiting Faculty
with Indian Institute of Technology (Bombay) and Global Foundries Visiting
Chair with National University of Singapore, Singapore. He has been in the editorial
board of IEEE Design and Test, IEEE TRANSACTIONS ON CIRCUITS AND
SYSTEMS, IEEE TRANSACTIONS ON VLSI SYSTEMS, and IEEE TRANSACTIONS
ON ELECTRON DEVICES. He was a Guest Editor for special issue on low-power
VLSI in the IEEE Design and Test (1994) and IEEE TRANSACTIONS ON VLSI
SYSTEMS (June 2000), IEE Proceedings—Computers and Digital Techniques
(July 2002), and IEEE JOURNAL ON EMERGING AND SELECTED TOPICS IN CIRCUITS
AND SYSTEMS (2011).
\end{IEEEbiography}

% that's all folks
\end{document}

%% file: sec/0-abstract.tex
\begin{abstract}
In this work, we propose stochastic Binary Spiking Neural Network (sBSNN) composed of stochastic spiking neurons and binary synapses (stochastic only during training) that computes probabilistically with one-bit precision for power-efficient and memory-compressed neuromorphic computing. We present an energy-efficient implementation of the proposed sBSNN using \textit{`stochastic bit'} as the core computational primitive to realize the stochastic neurons and synapses, which are fabricated in 90nm CMOS process, to achieve efficient on-chip training and inference for image recognition tasks. The measured data shows that the \textit{`stochastic bit'} can be programmed to mimic spiking neurons, and stochastic Spike Timing Dependent Plasticity (or sSTDP) rule for training the binary synaptic weights without expensive random number generators. Our results indicate that the proposed sBSNN realization offers possibility of up to 32$\times$ neuronal and synaptic memory compression compared to full precision (32-bit) SNN and energy efficiency of 89.49 TOPS/Watt for two-layer fully-connected SNN.
\end{abstract}


%% file: sec/1-introduction.tex
\section{Introduction}\label{sec:intro}

% Deep Learning
\IEEEPARstart{I}{n} the current era of ubiquitous autonomous intelligence, there is a growing need for moving Artificial Intelligence (AI) to the edge to cope with the ever increasing demand for autonomous systems like drones, self-driving cars, and smart wearable devices. Energy-efficient neuromorphic systems are henceforth necessary to process the massive amount of data generated by the resource-constrained battery-powered edge devices. Furthermore, it is highly desirable to embed on-chip intelligence using low-complexity learning rules, which enable the edge devices to learn from real-time inputs. Real-time on-chip learning capability precludes the need for offline training in the cloud, which can otherwise lead to higher latency and security concerns for real-time applications.

Spiking Neural Networks (SNNs), on the account of event-driven computing capability and hardware-friendly local learning using Spike Timing Dependent Plasticity (STDP), offer a promising solution for realizing energy-efficient neuromorphic systems with on-chip intelligence. In fact, researchers in \cite{blouw2018benchmarking} demonstrated that SNN running on event-driven neuromorphic hardware like Intel \textit{Loihi} \cite{davies2018loihi} incurs the minimum energy per inference relative to similarly sized analog neural network executed on CPU/GPU while providing equivalent inference accuracy for a latency-critical keyword spotting task. Recent works on deep SNNs indicate that energy efficiency significantly increases with network depth due to exponential drop in the spiking activity across successive SNN layers \cite{sengupta2019going, lee2019enabling}. In this regard, prior works proposed energy-efficient implementations of SNN using CMOS \cite{akopyan2015truenorth, davies2018loihi, cheung2016neuroflow} and emerging device technologies such as Resistive Random Access Memory (RRAM) \cite{linares2009memristance, milo2016demonstration}, Conductive Bridge RAM (CBRAM) \cite{shi2018neuroinspired}, and Magnetic Tunnel Junctions (MTJs) \cite{sengupta2016hybrid}. However, SNNs composed of deterministic neuronal and synaptic models require multi-bit precision to store the parameters governing their dynamics. As a result, the computational complexity and neuronal/synaptic memory requirements increase with network size, leading to reduction in the overall power- and area-efficiency.

We propose and implement \textit{`stochastic bits'} enabled binary SNN (sBSNN) that computes probabilistically with one-bit precision for energy- and memory-efficient neuromorphic computing at the edge. The core building block of the sBSNN is a \textit{`stochastic bit'}, which switches between its logic low and high states with a probability that varies in a sigmoidal manner based on the input. We realize the stochastic neurons, referred to as sNeurons, and synapses (stochastic only during training) using the proposed \textit{`stochastic bit'} as explained below. The sNeuron receives the weighted sum of the input spikes with the synaptic weights, and spikes probabilistically depending on the weighted input sum. The firing probability of the sNeurons, similar to the switching dynamics of the \textit{`stochastic bit'}, has sigmoidal relationship with the weighted input sum. The binary synapse interconnecting a pair of input (pre) and output (post) neurons is similarly emulated using the \textit{`stochastic bit'} during training. The binary synaptic weight is trained using the stochastic-STDP (sSTDP) algorithm presented in \cite{srinivasan2016magnetic}, where the synaptic weight is potentiated/depressed with a probability that depends on the degree of correlation between the spike times of the pre- and post-neurons. The trained binary synaptic weights are then used deterministically during inference to predict the class of a test input. The proposed sBSNN, with event-driven computing capability enabled by state-less sNeurons and memory-efficient on-chip learning capability enabled by the hardware-friendly localized sSTDP rule, offers a promising solution for building the next generation of autonomous intelligent systems.
\\

To that effect, we propose an energy-efficient realization of sBSNN, fabricated in 90nm CMOS technology, to achieve on-chip training and inference for visual image recognition tasks. The proposed \textit{`stochastic bit'} is composed of a cross-coupled inverter with PMOS header and NMOS footer transistors for obtaining the sigmoidal switching probability characteristics. We interface the CMOS \textit{`stochastic bit'} with the appropriate peripheral circuitry to realize the sNeurons and synapses. The energy and memory efficiency of the proposed implementation stems from three key factors. First, the power consumed by the sNeuron for generating a spike is comparable to that consumed in a single transition of a cross-coupled inverter, which is typically in the order of few $\mu W$. In addition, the \textit{`stochastic bit'} design also leverages power gating technique \cite{mutoh19951} with header and footer transistors between the supply and ground rails for reducing the leakage power consumption. Second, the spiking dynamics of the sNeuron depend only on the current input and not on the integrated sum of the current and past inputs, which precludes the need for storing the neuron state (typically known as the membrane potential) as is common in deterministic spiking neurons like the leaky integrate-and-fire neuron. Further, the synapses need only one-bit storage to record the respective binary states. Last, the weighted sum of the inputs with the synaptic weights, which is typically a series of multiply and accumulate (MAC) operations in analog neural networks, is transformed to AND operations followed by pulse count in the proposed sBSNN, thereby reducing the computational energy significantly. Our analysis using a two-layer fully-connected SNN of 400 neurons indicates that the proposed realization offers high energy efficiency of 89.49 TOPS/Watt, which renders it a potential candidate for enabling the next generation of intelligent devices. 

%\revised{
In summary, we make the following contributions:
\begin{itemize}
    \item We proposed the \textit{`stochastic bit'} as the core computational primitive to realize the stochastic neurons and binary synapses, which are implemented in 90nm CMOS process. 
    \item We proposed and evaluated the \textit{`stochastic bit'} enabled sBSNN that computes probabilistically with one-bit precision for power-efficient and  memory-compressed neuromorphic computing. 
    \item We proposed and demonstrated one of the first works on all-CMOS realization of stochastic SNNs. Our proposal provides reconfigurable on-chip learning that is suitable for the real-time and resource constrained edge devices. 
\end{itemize}
%}
%We note that prior works on hardware implementations of stochastic SNNs were based on emerging technologies like CBRAMs and MTJs \cite{suri2012cbram, srinivasan2016magnetic}. Our proposal is one of the first works on all-CMOS realization of stochastic SNNs.

The rest of the paper is organized as follows. Section II details the proposed sBSNN and sSTDP training rule. Section III describes the \textit{`stochastic bit'} circuit design and the required peripherals for implementing the sNeurons and synapses. The system-level implementation of the sBSNN is also detailed in this section. Section IV presents the measured characterization results of the sNeurons and synapses, and the accuracy and energy efficiency of our sBSNN realization. Finally, section V concludes the paper.

%% file: sec/2-background.tex
\section{Background}\label{sec:background}
\subsection{Stochastic Binary Spiking Neural Network (sBSNN)}\label{sec:background-sbsnn}
\begin{figure}[t]
  \centering
  \includegraphics[width=1.0\columnwidth]{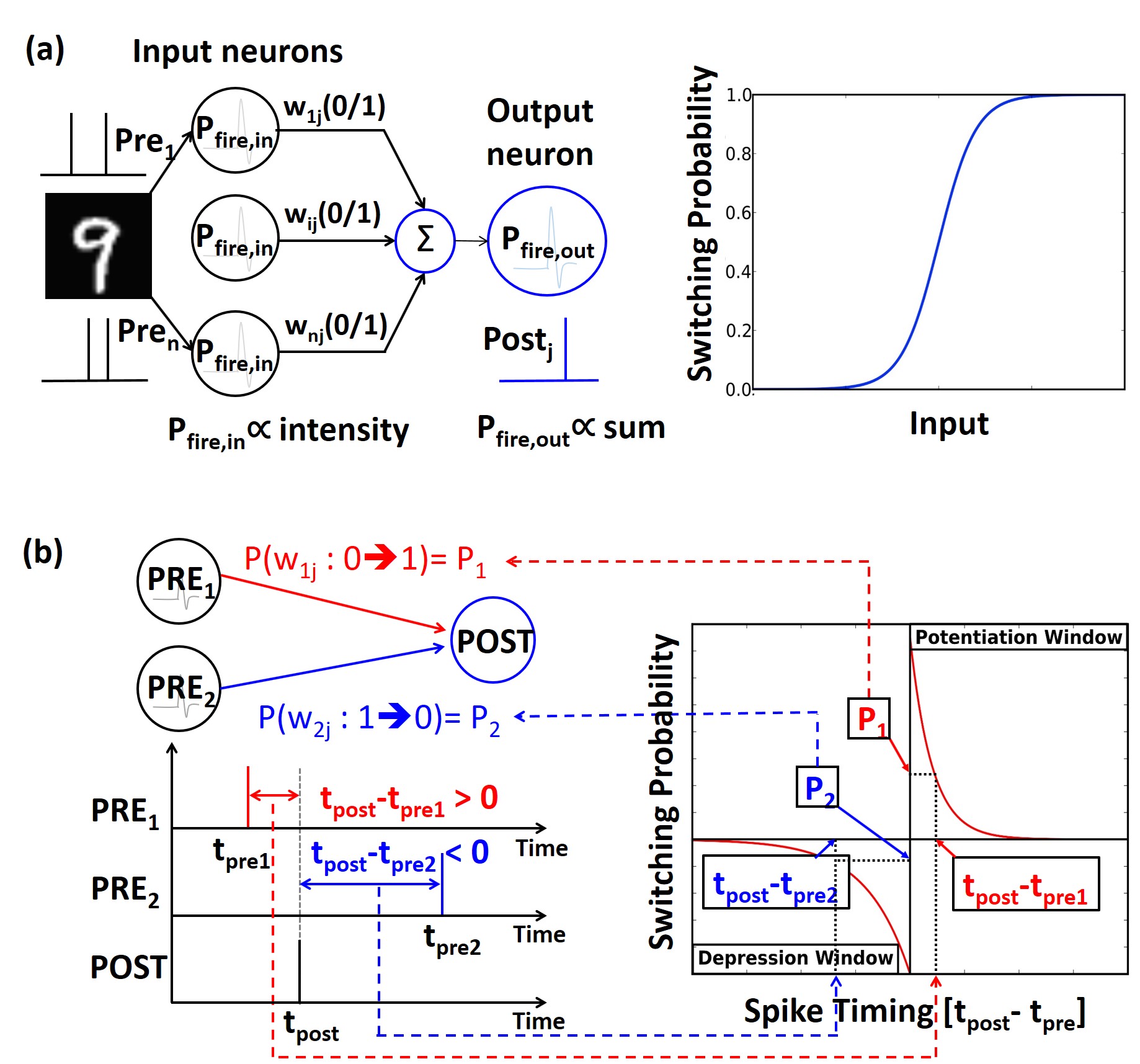}
  %\shrinkBeforeFigureCaption
  \caption{(a) SNN composed of stochastic input and output neurons interconnected via binary synaptic weights. The output sNeurons fire with a probability that has sigmoidal relationship with the weighted input sum. (b) Stochastic-STDP learning rule for binary synaptic weights, where the synaptic switching probability depends on the time difference between input (pre) and output (post) spikes. The binary weight is probabilistically potentiated (depressed) for positive (negative) timing correlation between pre- and post-spikes.
}\label{fig:SB-SNN}
  %\shrinkAfterFigure
\end{figure}

The core building block of the proposed sBSNN is a set of input (pre) neurons connected to an output (post) neuron via binary weights. The input neurons, which represent the image pixels for a visual object recognition task, generate Poisson-distributed spikes at a rate proportional to the corresponding pixel intensities. At any given time, the input pre-spikes get modulated by the interconnecting synaptic weights to produce resultant current into the output neuron. 
%\revised{
Several previous works have explored the hardware implementations for these core building blocks of stochastic SNNs, using emerging technologies like CBRAMs and MTJs \cite{suri2012cbram, srinivasan2016magnetic} and built-in blocks in FPGA board \cite{yousefzadeh2018practical}. 
We proposed a \textit{`stochastic bit'} as the core building block for neuron and synapse (training) to achieve on-chip learning with compressed memory.
%}
We model the output neuron using the \textit{`stochastic bit'}, which spikes probabilistically based on the input current (or weighted input sum) during both training and inference. The spiking probability of the output sNeuron has sigmoidal dependence on the input current as illustrated in Fig.~\ref{fig:SB-SNN}(a). It is important to note that the sNeuron is state-less since the stochastic spiking dynamics depend only on the instantaneous input current and not on the integrated sum of current and past input currents as is typical in deterministic neuron models, thereby eliminating the multi-bit precision requirement for the neuron state (or membrane potential). The stochastic synapses (stochastic only during training) are similarly emulated using the \textit{`stochastic bit'}, where the synaptic switching probability depends on the time difference between the pre- and post-spikes as explained in the following \autoref{sec:StocSTDP}.

\subsection{Stochastic-STDP (sSTDP)}\label{sec:StocSTDP}
Spike Timing Dependent Plasticity (STDP) is a bio-inspired local learning mechansim, which has been experimentally observed in the rat hippocampus ~\cite{bi1998synaptic}. STDP postulates that the change in the weight of a multi-level synapse interconnecting a pair of pre- and post-neurons depends on the correlation between the respective spike times. 
If the pre-neuron spikes before the post-neuron, the synaptic weight increases (synaptic potentiation), while it decreases if the pre-neuron spikes after the post-neuron (synaptic depression). Binary synapses, on the contrary, require a probabilistic learning rule to prevent rapid switching of the weights between the high and low levels, which would otherwise render the synapses memory-less. We use the sSTDP learning algorithm proposed in ~\cite{srinivasan2016magnetic} to train the binary synaptic weights, where the synaptic switching probability has exponential dependence on spike timing difference as illustrated in Fig. \ref{fig:SB-SNN}(b) and described by
\begin{eqnarray} \label{eqn:StoCSTDP}
 P_{L \xrightarrow{} H} = \gamma_{pot} \cdot e^{\frac{-\Delta t}{\tau_{pot}}} where
 \ \Delta t = t_{post} - t_{pre} > 0 \\
 P_{H \xrightarrow{} L} = \gamma_{dep} \cdot e^{\frac{\Delta t}{\tau_{dep}}} where \ \Delta t = t_{post} - t_{pre} < 0
\end{eqnarray}
where $P_{L \xrightarrow{} H}$ and $P_{H \xrightarrow{} L}$ are the probability of potentiation and depression, respectively. 
%\revised{
In other words, the weight of a synapse changes based on the temporal correlation between the spike time of pre- and post-neurons.
For example, if a pre- (post-) neuron fires before a post- (pre-) neuron does, it is positively (negatively) correlated with the input pattern \cite{lowel1992selection}. Consequently, potentiation (depression) occurs probabilistically in the positive (negative) timing window of the sSTDP algorithm. The corresponding switching probability is determined by the spike timing difference between pre and post spikes as described in the above equations. 
%}
The peak switching probability and time constant for potentiation ($\gamma_{pot},\tau_{pot}$) and depression ($\gamma_{dep},\tau_{dep}$) determine the synaptic learning efficacy. The sSTDP hyperparameters have to be chosen carefully to ensure right balance between the potentiation and depression weight updates, and achieve efficient learning. Once the training is complete, the learnt binary weights are used deterministically during inference. The presented sBSNN requires only one-bit precision for the neurons and synapses, leading to visual image recognition with compressed memory requirement.

%% file: sec/3-stochastic-bit.tex
\begin{figure}[t]
  \centering
  \includegraphics[width=1.0\columnwidth]{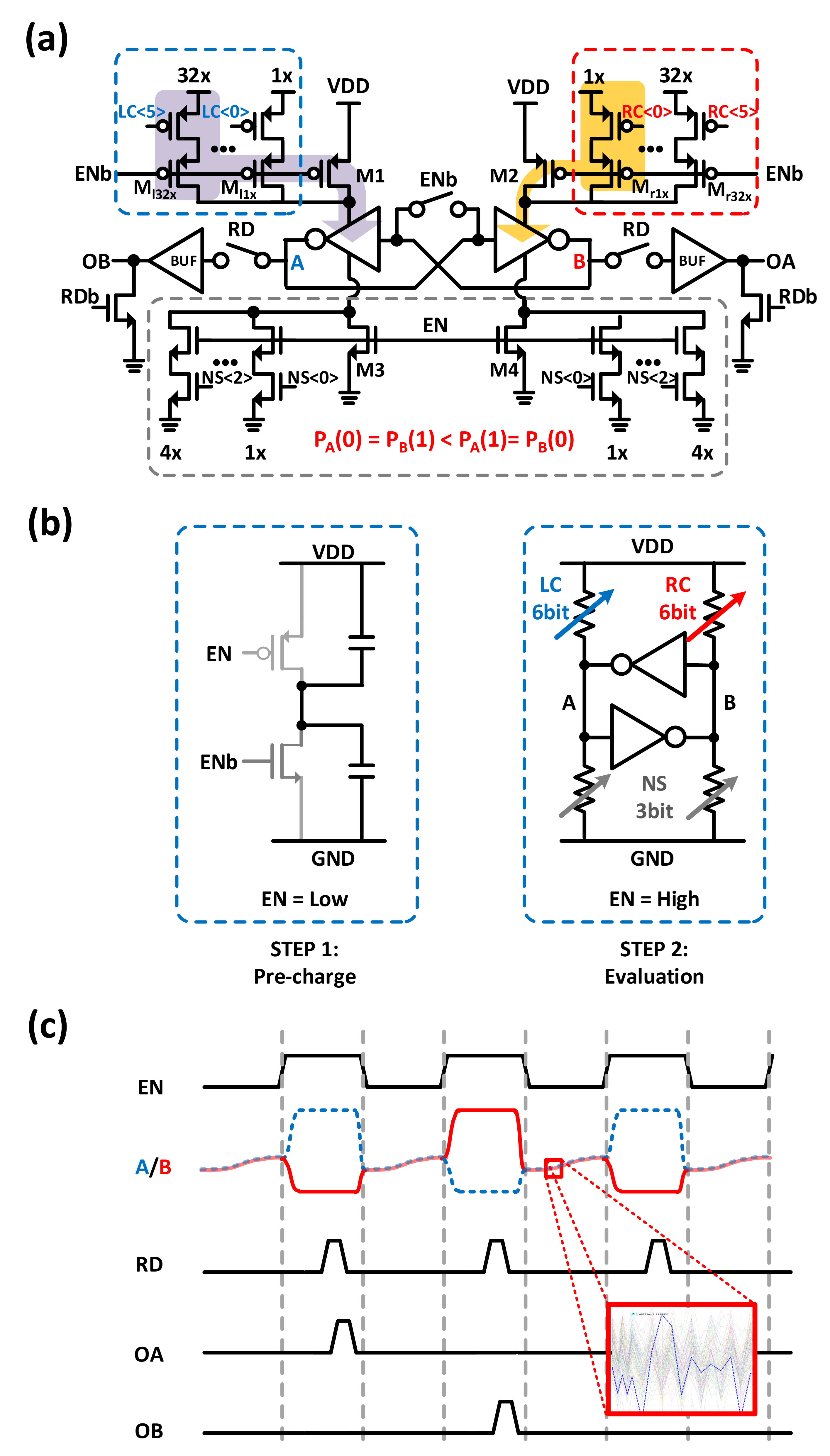}
  %\shrinkBeforeFigureCaption
  \caption{(a) Schematic of 6-bit \textit{`stochastic bit'} core. (b) Illustration of the pre-charge and evaluation modes of operation of the \textit{`stochastic bit'}. (c) Timing diagram illustrating the operation of the \textit{`stochastic bit'}.
}\label{fig:stochastic-bit}
  %\shrinkAfterFigure
\end{figure}

\section{sBSNN Design and Implementation}\label{sec:sto-bit}
In this section, we first detail the design and implementation of the proposed \textit{`stochastic bit'}, which is the core computing primitive of the sBSNN. We then present the design of sNeuron and synapse (stochastic only during training). Finally, we detail the system-level realization of two-layer fully-connected sBSNN for visual image recognition.

\subsection {CMOS `Stochastic bit' Design} \label{sec:sto_circuit_design}
As mentioned in \autoref{sec:intro}, controllable stochastic behavior is the central characteristic of the \textit{`stochastic bit'}. In CMOS-based designs, stochastic behavior is largely dependent on the characteristics of the random noise source. Thermal noise is one of the commonly used entropy sources in CMOS process, which stems from the channel fluctuations induced by random Brownian motion of electrons. The power spectral density of thermal noise across a resistor is given by $V^{2} = 4kTR$, where $k$ is the Boltzman constant, $T$ is the temperature in Kelvin, and $R$ is the resistance in ohms. Accordingly, thermal noise induced stochasticity is only affected by the device resistance and operating temperature. Thermal noise has been used as the source of randomness in many True Random Number Generator (TRNG) designs \cite{holleman20083,petrie2000noise}. Also, metastability-based TRNG designs using cross-coupled inverters have been reported to achieve high operating frequency and power efficiency \cite{mathew20122}. This motivated us to investigate the possibility of harnessing the metastable behavior of bi-stable circuits to implement the \textit{`stochastic bit'}.

The proposed \textit{`stochastic bit'} is realized using cross-coupled inverter with PMOS header transistors and NMOS footer transistors as depicted in Fig.~\ref{fig:stochastic-bit}(a). The operation of the \textit{`stochastic bit'} is divided into two different modes, namely, pre-charge and evaluation, which are gated by the `EN' (enable) signal as shown in Fig.~\ref{fig:stochastic-bit}(b). In the pre-charge mode (when `EN' is low), the cross-coupled nodes A and B are pre-charged to the same voltage by leakage current, while the header and the footer transistors are turned off. Note that, the inherent power gating enabled by the PMOS header transistors and the NMOS footer transistors causes the leakage current of the proposed design to be lower than the gate leakage current of a 6T SRAM bitcell \cite{mutoh19951}. 
%Once `EN' is driven high for evaluation, one of the nodes A or B probabilistically pulls up to supply voltage under the influence of thermal noise while the other node pulls down to ground. 
The switching probability depends on asymmetry in the effective strength of left- and right-wing PMOS transistors, which can be modulated using the input that is represented as 6-bit code in our implementation and activates different binary weighted PMOS switch transistors. 
%For example, if the 32$\times$ weighted left PMOS transistor and 1$\times$ weighted right PMOS transistor are the only header transistors that are turned on as highlighted in Fig.~\ref{fig:stochastic-bit}(a), node A is more likely pulled up to supply voltage due to the stronger current driving capability of the 32$\times$ PMOS transistor than that of 1$\times$ PMOS transistor. As a result, the probability of logic high state at node B ($P_{B}(1)$), the output node of the left inverter, is higher than that at node A ($P_{A}(1)$). 
The NMOS footer transistors connected to ground are controlled symmetrically in strength using the same input code, which is represented with 3-bit precision in our implementation, to modulate the shape of the probability curve. 
%Note that the PMOS and NMOS codes are set during the pre-charge phase, and the output of nodes A and B are buffered to nodes OA and OB when the read signal (RD) is asserted high as shown in Fig.~\ref{fig:stochastic-bit}(c). 
%The buffered output nodes are driven to ground by default, which causes only one of the nodes OA or OB to generate an output pulse. 
The shape of the switching probability versus the PMOS digital code is sigmoidal as will be shown in the results \autoref{sec:result}. The shape and the covered range of probability is programmable and can be reconfigured on-chip. 
It is worth noting that, the \textit{`stochastic bit'} consumes only leakage power during the pre-charge mode, and charging/discharging power for nodes A and B during the evaluation mode. In addition, the speed of operation is based on the speed of `EN' signal. Therefore, the proposed design becomes more power efficient and faster as CMOS process scales. Also, the PMOS and NMOS sizing, and bit-precision for the respective codes can be tuned based on the application requirements.

\begin{figure}[t]
  \centering
  \includegraphics[width=0.8\columnwidth]{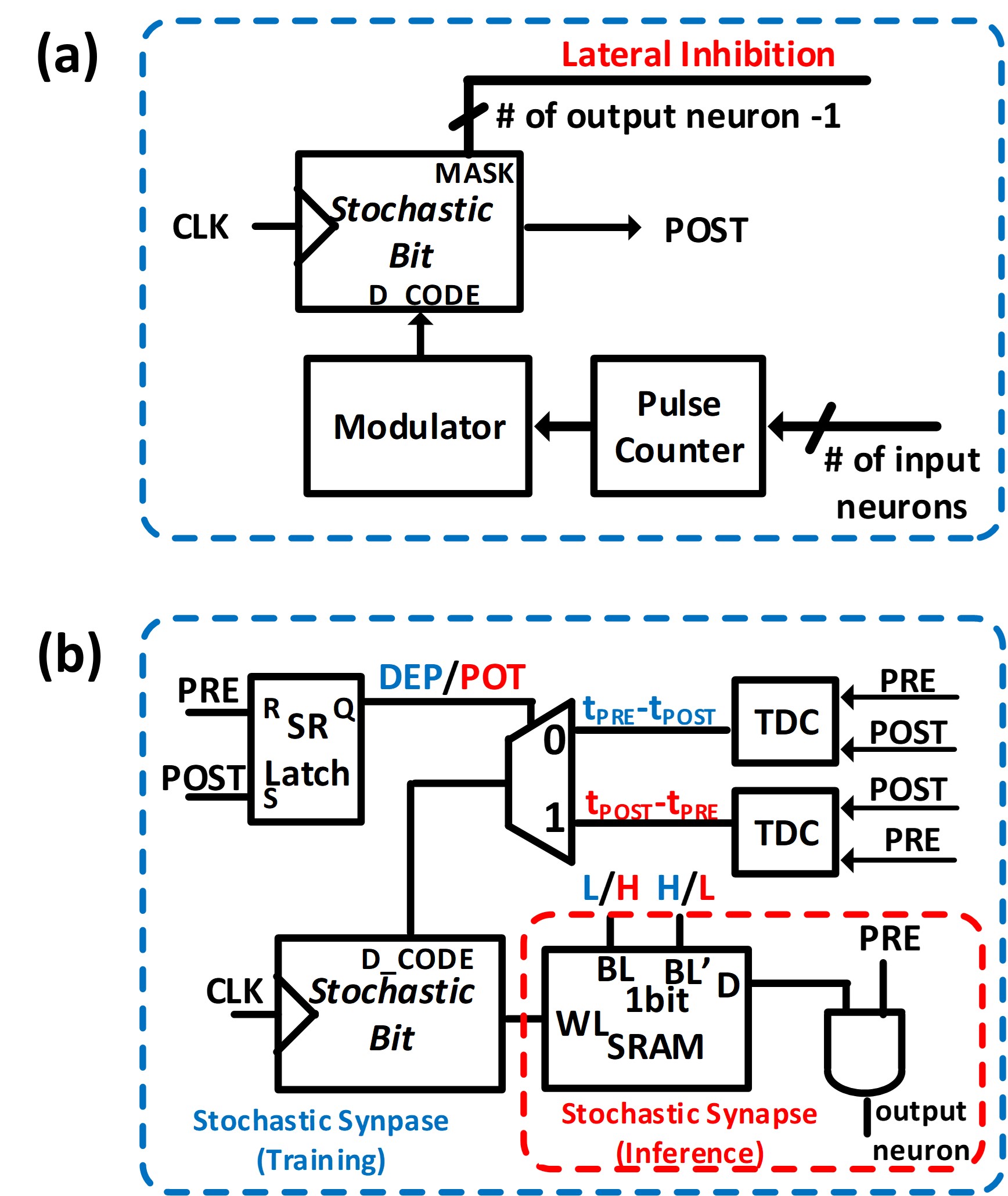}
  %\shrinkBeforeFigureCaption
  \caption{Design of `Stochastic bit' enabled (a) spiking neuron, and (b) binary synapse (stochastic during training and deterministic during inference).
}\label{fig:stochastic-neurons&synapses}
  %\shrinkAfterFigure
\end{figure}

\subsection {Stochastic Neuron (sNeuron)}\label{sec:sto-neuron}
We now describe how the \textit{`stochastic bit'} is used to realize stochastic input and output neurons forming the sBSNN. The input neurons map the image pixel intensities to spike trains, where each neuron fires probabilistically at a rate proportional to the corresponding pixel intensity. The \textit{`stochastic bit'} can inherently realize an input sNeuron by mapping the pixel intensity to PMOS code that controls its switching probability. On the contrary, the \textit{`stochastic bit'} is interfaced with counter and modulator circuit (shown in Fig.~\ref{fig:stochastic-neurons&synapses}(a)), which generates and modulates the weighted input, for realizing the output sNeuron that spikes with the desired probability. 
Also, the spiking activity of the sNeuron can be suppressed by masking the `EN' signal of the \textit{`stochastic bit'}, which is used for implementing lateral inhibition that facilitates competitive learning as will be explained in \autoref{sec:sto-system}.
%\revised{
The generated spikes from the input and the output sNeuron (PRE and POST) are applied to the stochastic binary synapses for synaptic updates as explained below.  
%}
%The neuronal firing probability range and dispersion can be programmed on-chip using the input codes for the \textit{`stochastic bit'} and the modulator.

\subsection {Stochastic Binary Synapse}\label{sec:sto-synapse}
The stochastic binary synapse (during training) is realized by interfacing the \textit{`stochastic bit'} with 6T SRAM as depicted in Fig.~\ref{fig:stochastic-neurons&synapses}(b). Based on the sign of the spike timing difference, $t_{post} - t_{pre}$, the synaptic weight update event is determined as potentiation (depression) when the sign is positive (negative). Then, the spike timing difference, measured as the number of clock pulses 
using time to digital converter (TDC), feeds the \textit{`stochastic bit'} to selectively turn on the PMOS header transistors, effectively causing it to produce an output pulse with the appropriate probability depending on spike timing. 
%\revised{
Note that TDC can be realized using a counter for potentiation (depression) that resets when PRE (POST) is high. TDC is shared by stochastic synapses that are activated by the same PRE/POST signal.
%}
The generated pulse activates the wordline of the 6T SRAM cell while the bitline is driven to VDD (ground) for synaptic potentiation (depression) update. Once the stochastic training process is complete, the \textit{`stochastic bit'} is powered off and the learnt binary weight stored in the corresponding SRAM cell is deterministically used during inference as shown in \autoref{fig:stochastic-neurons&synapses}(b). Note that, during both training and inference, the computation of the weighted input sum reduces to AND operations followed by pulse count since both the inputs and synaptic weights are binary. Hence, the sBSNN provides much higher computational energy efficiency relative to analog neural networks with real-valued (32-bit) inputs and synaptic weights, which require MAC (multiply-and-accumulate) units, and SNNs with real-valued weights and binary inputs that need accumulators for computing the weighted input sum.
%The operation of stochastic synapse gets even simpler during inference as shown in Fig.~\ref{fig:stochastic-neurons&synapses}(b). The binary weighted synaptic value is only propagated to the membrane potential of the connected stochastic output neuron when the stored value in the 1-bit SRAM is `1' and there is a pre-spike. Note that, the stoc-STDP dynamics can also be programmed in a similar way that we described in the previous section for the stochastic neuron. Moreover, we can set unique P-STDP curves for different synapses, which facilitates programmable on-chip learning.

\begin{figure}[!b]
  \centering
  \includegraphics[width=0.75\columnwidth]{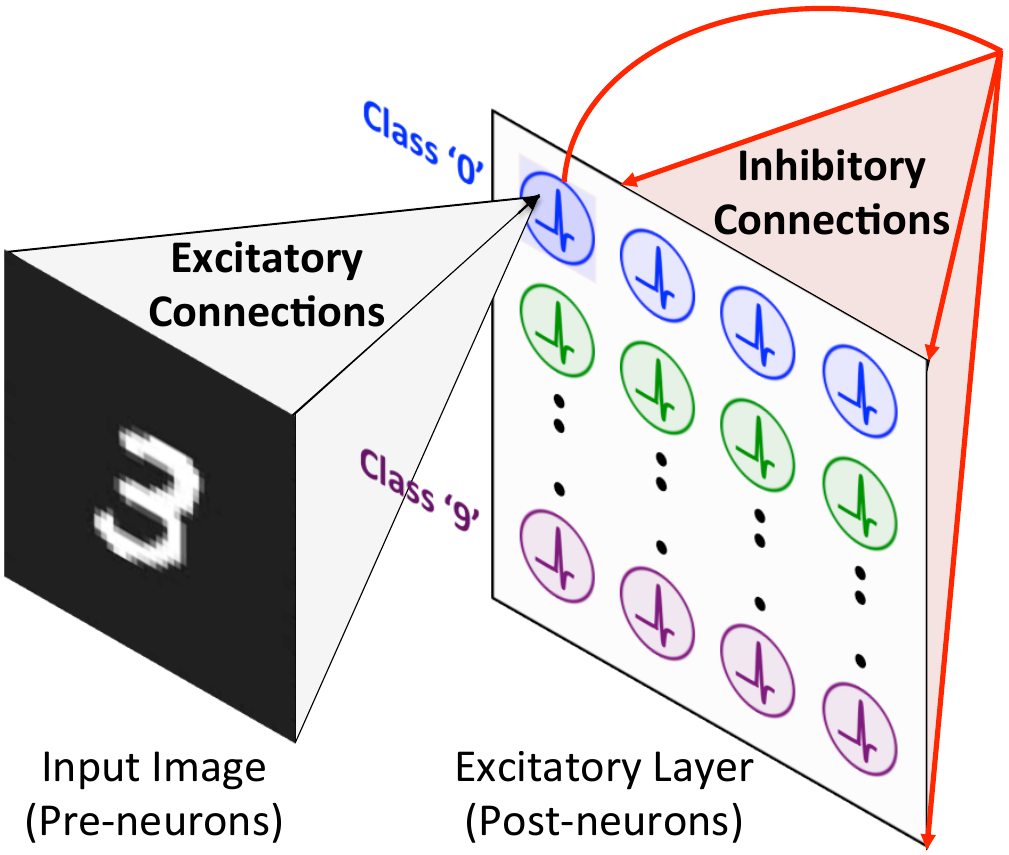}
  %\shrinkBeforeFigureCaption
  \caption{Architecture of two-layer fully-connected sBSNN, with lateral inhibition, for object recognition. The input pixels (pre-neurons) are fully-connected via stochastic binary synapses to excitatory post-neurons, which are trained using the sSTDP learning algorithm. The excitatory neurons are subdivided into different groups, where each group is trained on a unique class of image patterns.
}\label{fig:SBSNN_Arch}
  %\shrinkAfterFigure
\end{figure}

\begin{figure}[!b]
  \centering
  \includegraphics[width=1\columnwidth]{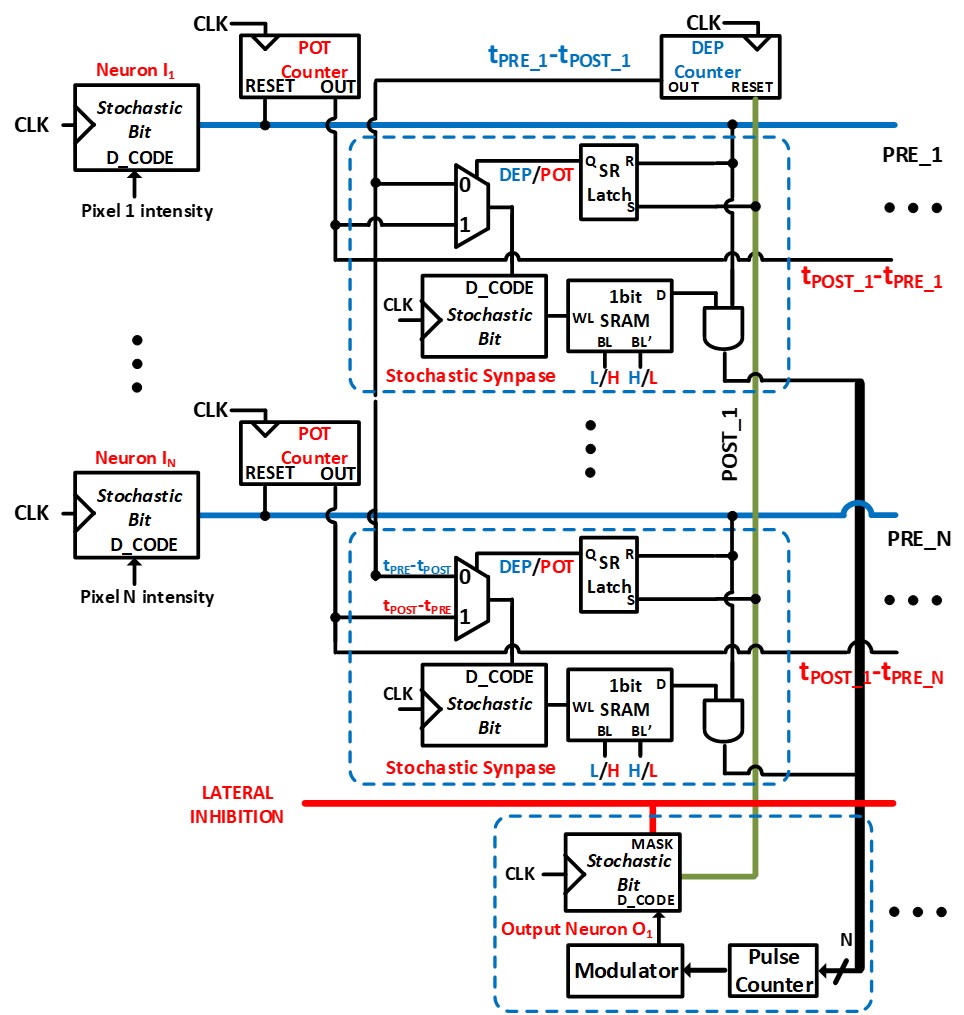}
  %\shrinkBeforeFigureCaption
  \caption{System-level realization of two-layer fully-connected sBSNN with lateral inhibition.
}\label{fig:implementation_SB-SNN}
  %\shrinkAfterFigure
\end{figure}

\begin{figure}[!t]
  \centering
  \includegraphics[width=0.9\columnwidth]{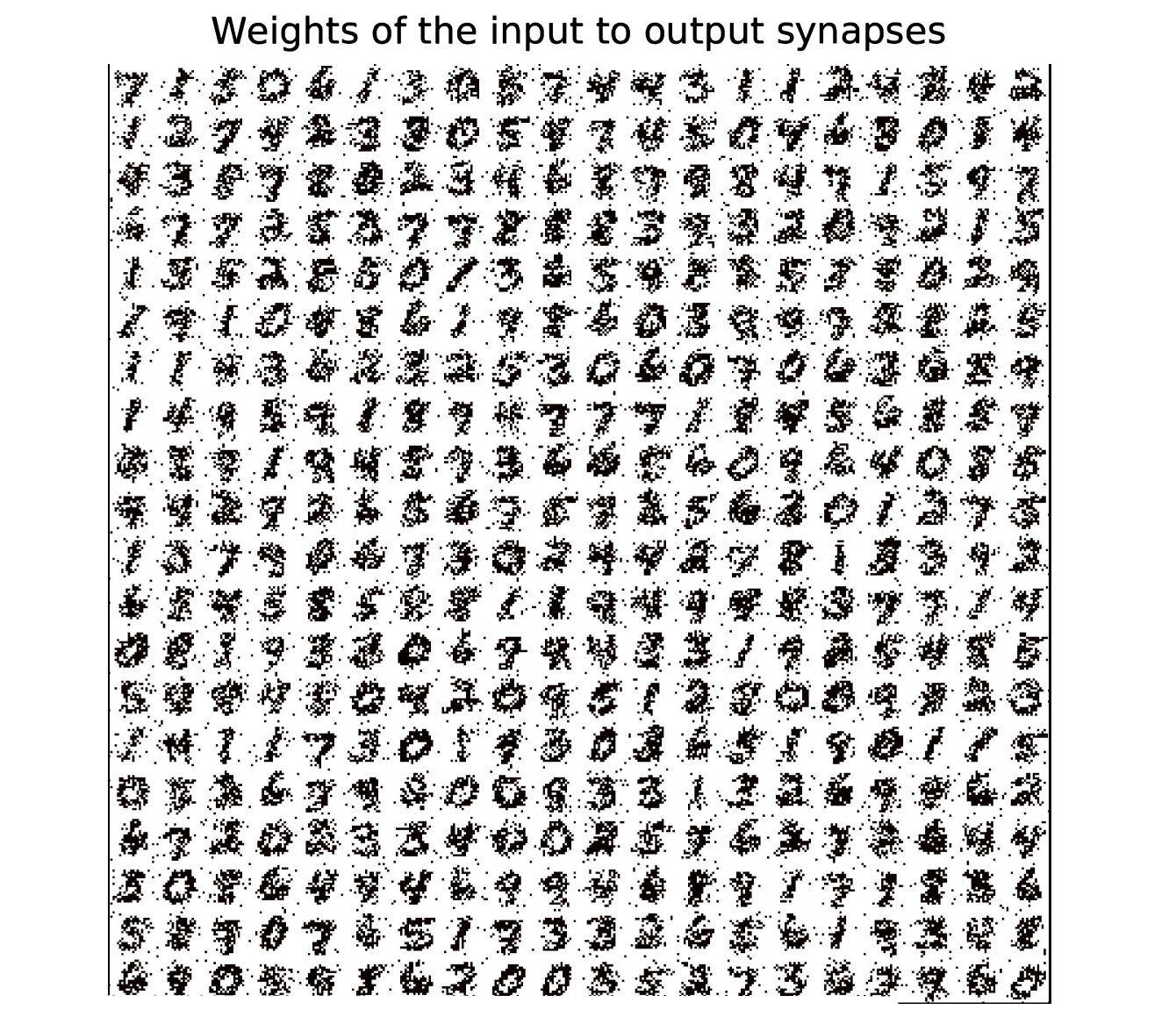}
  %\shrinkBeforeFigureCaption
  \caption{MNIST digit representations (28$\times$28 in dimension) learnt by a two-layer fully-connected sBSNN of 400 excitatory neurons (organized in 20$\times$20 grid).
}\label{fig:SBSNN_Digits}
  %\shrinkAfterFigure
\end{figure}

\subsection {sBSNN System-level Implementation}\label{sec:sto-system}
\textbf{On-chip training:} We demonstrate the efficacy of the proposed sNeuron and synapse using a two-layer fully-connected sBSNN depicted in Fig.~\ref{fig:SBSNN_Arch}. Fig. \ref{fig:implementation_SB-SNN} illustrates the system-level implementation of the two-layer sBSNN. The input sNeurons representing the image pixels are fully-connected via binary weights to output (post) sNeurons. At every time-step, the weighted sum of the input spikes with the synaptic weights are modulated and fed to the \textit{`stochastic bit'} in the respective post-neurons, causing them to fire probabilistically. The weighted sum received by each post-neuron is calculated by counting the number of pulses from the output of the AND gates in the corresponding column of synapses as depicted in Fig.~\ref{fig:implementation_SB-SNN}. The pulses are only generated when both inputs of the AND gate are high. Accordingly, power is only dissipated when there are transitions in the AND gate. As a result, the weighted input sum computation in sBSNN consumes significantly lower power compared to full precision (32-bit) SNN. In the event of a post-spike, the spiking neuron inhibits the remaining post-neurons, as illustrated in Fig.\ref{fig:SBSNN_Arch}, by masking the respective enable (EN) inputs as explained in \autoref{sec:sto-neuron} to uniquely learn the presented pattern. The synapses connecting the input to the spiking post-neuron are probabilistically potentiated based on spike timing. The spike timing difference, $t_{post} - t_{pre}$ ($t_{pre} - t_{post}$) in the number of clock pulses, is measured using the POT (DEP) counter shown in Fig. \ref{fig:implementation_SB-SNN}, which is reset at every pre-spike (post-spike) and decremented by unity at successive time-steps. The elapsed count of POT (DEP) counter is sampled upon a post-spike (pre-spike) for potentiation (depression) weight update. The spike timing difference is fed to the \textit{`stochastic bit'} in the synapses (depicted in Fig. \ref{fig:stochastic-neurons&synapses}(b)), which in turn probabilistically programs the SRAM as detailed in \autoref{sec:sto-synapse}. The sSTDP-based probabilistic weight updates enable each excitatory neuron to learn a complete representation of the input pattern in the input to excitatory synaptic weights. In order to ensure that each excitatory neuron learns unique input representations, we divided the excitatory neurons into different clusters and trained each cluster of neurons on a distinct class of input patterns as proposed in \cite{srinivasan2016magnetic}. Fig. \ref{fig:SBSNN_Digits} shows the MNIST digit representations learnt by a two-layer fully-connected sBSNN of 400 excitatory neurons using the sSTDP-based training methodology.

\textbf{On-chip inference:} At the end of training, each post-neuron learns to spike for a unique input class by encoding a general input representation in the input to output synaptic weights as shown in Fig. \ref{fig:SBSNN_Digits}. 
%\revised{
Once training is completed, we disable the clock signal of the \textit{`stochastic bit'} in the synapses, thereby fixing the weights for the inference phase.
%}
The learnt binary weights, stored in the SRAM cells, are used deterministically during inference. A test pattern is predicted to belong to the class learnt by the group of neurons with the highest average spike count over the time period for which the test input is presented. The proposed sBSNN implementation, by virtue of using simpler weighted input sum computation and state-less stochastic neurons, can provide high energy efficiency during inference as will be shown in \autoref{sec:result}.
%The SB-SNN is implemented with core computational primitive called the \textit{`stochastic bit'}, which is programmable and reconfigurable on-chip. Our proposed SB-SNN appreciates event-driven and stochastic computing, which consumes low power in both learning and inference. Also, stoc-STDP learning rule enables SB-SNN to compute MAC operation in a simple way with the stochastic binary synapses, leading to energy-efficient computation with compressed memory size. Furthermore, the \textit{`stochastic bit'}, which is a core computing primitive of SB-SNN, is implemented in CMOS process, keeping the benefits of scaling in power consumption and speed of computation.

%% file: sec/4-result.tex
\section{Results}\label{sec:result}
In this section, we first present the measured results of the sNeuron and synapse, which are fabricated in 90nm CMOS process. We subsequently show the simulation results of our sBSNN implementation (detailed in \autoref{sec:sto-system}) using the measured neuronal and synaptic dynamics on the MNIST dataset.

\begin{figure}[t]
  \centering
  \includegraphics[width=\columnwidth]{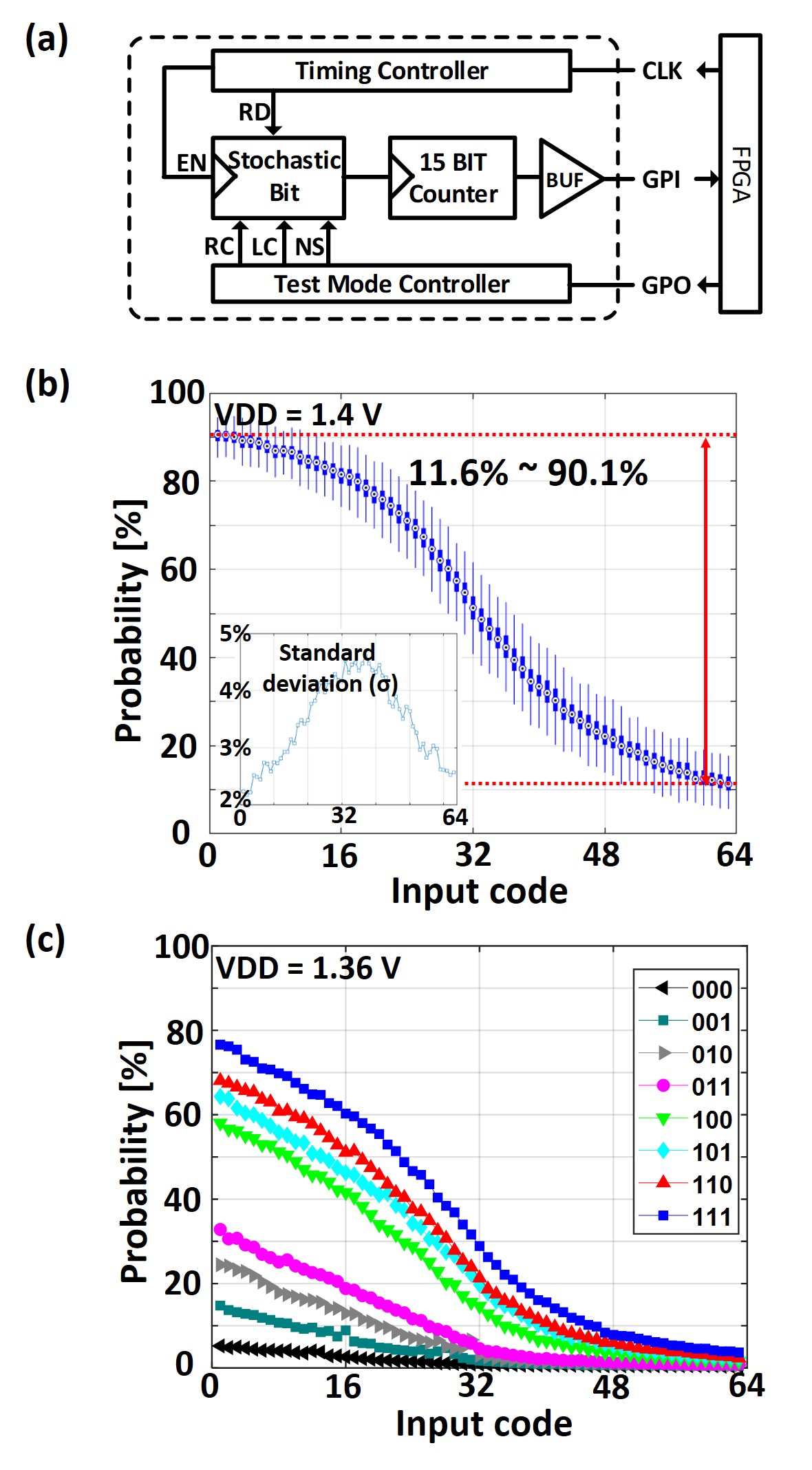}
  %\shrinkBeforeFigureCaption
  \caption{(a) Measurement setup for the \textit{`stochastic bit'} design, which is interfaced with a FPGA board for generating the on-chip clock/inputs and monitoring the outputs via the CLK and GPIO ports, respectively. (b) The measured box plots of switching probability versus the input (PMOS) digital code, and its standard deviation, $\sigma$ (refer to the inset). In each box, the central mark indicates the median, the ends of the vertical blue boxes indicate the 25th and 75th percentiles, respectively, and the lines indicate the min and max value. (c) Switching probability dynamics for different 3-bit NMOS codes. }\label{fig:measure_stochastic-bit}
  %\shrinkAfterFigure
\end{figure}

\subsection{`Stochastic bit' Characterization}\label{sec:result_sto}
Fig. \ref{fig:measure_stochastic-bit}(a) illustrates the setup for characterizing the CMOS \textit{`stochastic bit'} design (detailed in \autoref{sec:sto-bit}). The on-chip timing controller generates sufficient number of enable (EN) pulses, which is set to 768 in our experiments, for obtaining reasonable estimate of the \textit{`stochastic bit'} switching probability for a specific configuration of PMOS and NMOS codes. The number of resultant output pulses at OA (refer to Fig. \ref{fig:stochastic-bit}(a)) is recorded by a 15-bit on-chip counter to determine the switching probability for the chosen PMOS and NMOS codes. 
For every set of input codes, we performed the switching probability measurement 1000 times. Fig. \ref{fig:measure_stochastic-bit}(b) shows that the switching probability of the \textit{`stochastic bit'} varies roughly in a sigmoidal manner with the PMOS code. %On each box, the central mark shows the median, the ends of the vertical blue boxes indicate the 25th and 75th percentiles, respectively, and the lines indicate the min and max values. 
The measured switching probability ranges from $11.6\%$ to  $90.1\%$ with less than $5\%$ standard deviation at a supply voltage of 1.4$V$. In addition, we varied the NMOS code and found that it controls the shape of the switching probability curve as illustrated in Fig. \ref{fig:measure_stochastic-bit}(c). The variation in the switching probability dynamics with the NMOS code can be attributed to the change in the respective transistor sizes relative to the PMOS transistor sizes. Note that, the ratio of minimum to maximum switching probability is determined by the bit-precision of the PMOS code and the relative sizing (widths) of the PMOS and NMOS transistors, which need to be fixed at design-time based on the application requirements.

\begin{figure}[t]
  \centering
  \includegraphics[width=\columnwidth]{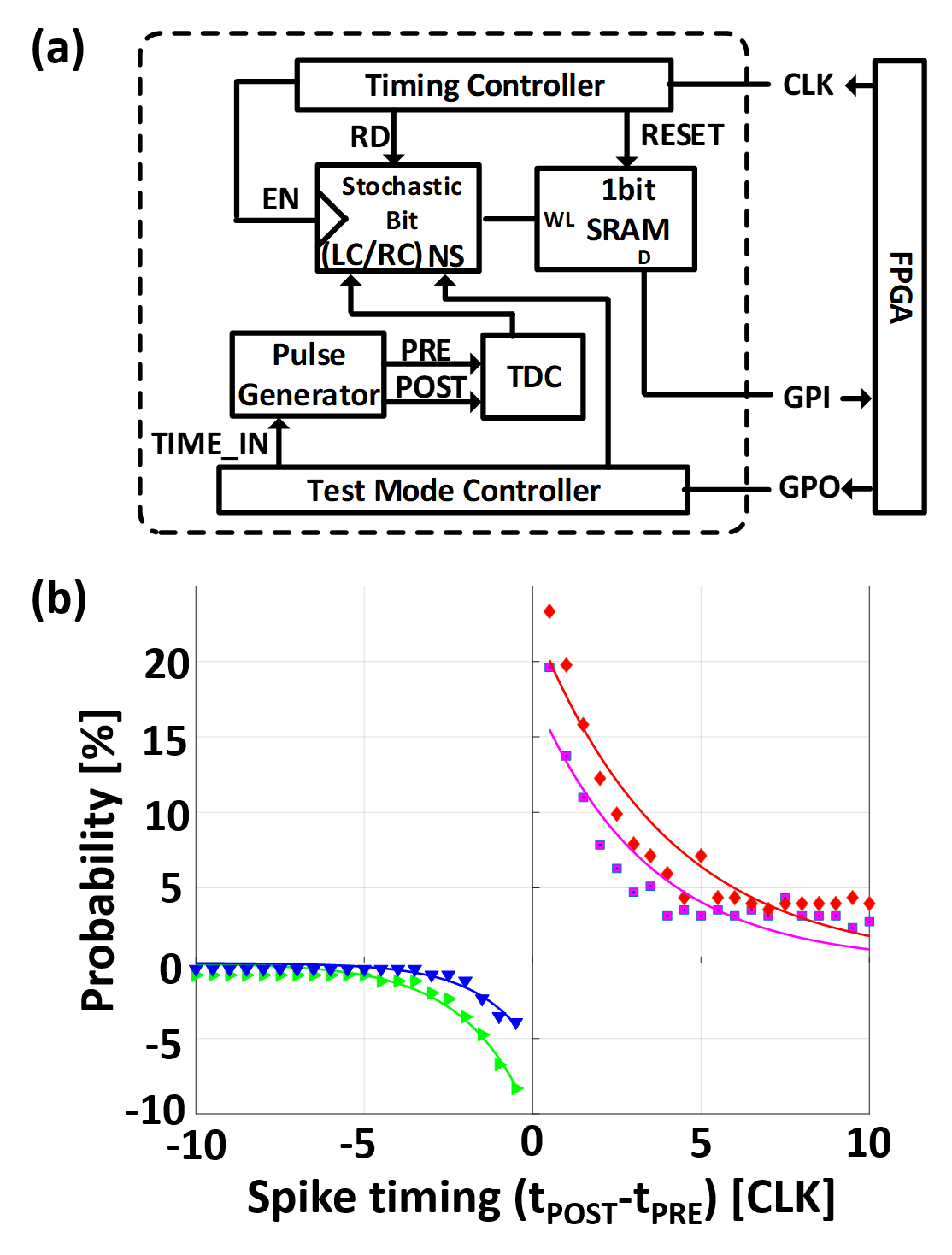}
  %\shrinkBeforeFigureCaption
  \caption{(a) Measurement setup for the sSTDP dynamics needed to train a stochastic binary synapse, which is interfaced with an FPGA board for generating the clock and inputs (spike timing, TIME\_IN), and monitoring the outputs (state of SRAM cell). (b) The measured sSTDP curve for different NMOS codes.}\label{fig:measure_pstdp}
  %\shrinkAfterFigure
\end{figure}

\subsection{Stochastic Binary Synapse}\label{sec:result_synapse}
The sSTDP dynamics required for training a binary synaptic weight are obtained by feeding the spike timing difference to the on-chip pulse generator, which generates the pre- and post-spikes as shown in Fig. \ref{fig:measure_pstdp}(a). The Time-to-Digital Converter (TDC) measures the spike timing difference and produces the PMOS code for the \textit{`stochastic bit'}, which probabilistically activates the SRAM wordline. The SRAM cell is then probed for potentiation (depression) event to estimate the sSTDP characteristics for the positive (negative) timing window. We adopted a methodology similar to that used for the \textit{`stochastic bit'} characterization for measuring the sSTDP dynamics as explained below. For every value of spike timing within the sSTDP window, TIME\_IN in Fig. \ref{fig:measure_pstdp}(a), we generated sufficient number of enable pulses (set to 768 as explained in \autoref{sec:result_sto}) for the \textit{`stochastic bit'} constituting the binary synapse. We then probed the 6T SRAM for a change in the cell state to determine the corresponding switching probability. We repeated the switching probability measurement 1000 times for every value of spike timing. Fig. \ref{fig:measure_pstdp}(b) shows the measured sSTDP dynamics, where the synaptic switching probability has roughly exponential dependence on spike timing, which conforms to the sSTDP rule depicted in Fig. \ref{fig:SB-SNN}(b). The sSTDP dynamics can be tuned on-chip by programming the NMOS code controlling the footer transistor sizes in the \textit{`stochastic bit'} as explained in \autoref{sec:sto_circuit_design}. Note that the Time-to-Digital Converter (TDC in Fig. \ref{fig:measure_pstdp}(a)) and pulse generators are used only for measurements. The binary synapse is composed of only the 6T-SRAM and the \textit{`stochastic bit'} during training, where the pre- and post-spikes are generated by the input and output sNeurons, respectively, constituting the sBSNN. Also, the spike timing difference is estimated using a counter per pre-/post-neuron as described in \autoref{sec:sto-system}. %For over 1000 iterations of `EN' pulses for different spike timing difference, `TIME\_IN', 1-bit stored in 6T SRAM is measured to check the bit flipping. The stoc-STDP dynamics can be modulated by changing probability of the \textit{`stochastic bit'}. Fig.~\ref{fig:measure_pstdp} (b) shows 4 different cases of P-STDP curve with different NMOS code, confirming tunable on-chip learning. The P-STDP curves used for the system-level simulations of SB-SNN are described in the following section.

\begin{table}[]
\Large
%\centering
\caption{Comparison with related works.}
\resizebox{\columnwidth}{!}{
\label{tab:comparison}
\begin{tabular}{|c|c|c|c|c|c|}
\hline
& \textbf{This Work}                                                                 & \textbf{2016 VLSI~\cite{ambrogio2016novel}}                                                              & \textbf{\begin{tabular}[c]{@{}c@{}}2015 \\ IEDM~\cite{kim2015nvm}\end{tabular}}    & \textbf{\begin{tabular}[c]{@{}c@{}}2017 VLSI \\ Report~\cite{jerry2017ultra}\end{tabular}} & \textbf{\begin{tabular}[c]{@{}c@{}}2015 \\ TCAS II~\cite{wu2015cmos}\end{tabular}} \\ \hline

\textbf{\begin{tabular}[c]{@{}c@{}}Learning\\ Rules\end{tabular}}             & \begin{tabular}[c]{@{}c@{}}Stochastic\\ STDP\end{tabular}                       & \begin{tabular}[c]{@{}c@{}}Stochastic\\ STDP\end{tabular}                       & STDP                                                             & \begin{tabular}[c]{@{}c@{}}Modulated\\ STDP~\cite{neftci2014event}\end{tabular}
                                   & STDP                                                             \\ \hline
\textbf{\begin{tabular}[c]{@{}c@{}}STDP\\ Timing Window\end{tabular}}         & \begin{tabular}[c]{@{}c@{}}267ns\\ (10 time steps)\\ @37.5MHz\end{tabular}         & 10ms                                                                            & 100us                                                            & N/A                                                                       & 3.5us                                                            \\ \hline
\textbf{\begin{tabular}[c]{@{}c@{}}Stochastic \\ deviation\end{tabular}} & \textless{}5\%                                                                      & N/A                                                                    & N/A                                                     & N/A                                                               & N/A                                                     \\ \hline
\textbf{\begin{tabular}[c]{@{}c@{}}On-chip\\ reconfigurable\end{tabular}}     & YES                                                                                & N/A                                                                    & YES                                                              & YES                                                               & YES                                                              \\ \hline
\textbf{\begin{tabular}[c]{@{}c@{}}Energy\\ /spike/neuron\end{tabular}}       & \begin{tabular}[c]{@{}c@{}}8.4pJ*\\ /1.84pJ**\end{tabular}                         & N/A                                                                             & N/A                                                              & 11.9 $\mu W$***                                                                        & \begin{tabular}[c]{@{}c@{}}9.3pJ\\ /3.6pJ***\end{tabular}        \\ \hline
\textbf{\begin{tabular}[c]{@{}c@{}}System\\ Configuration\end{tabular}}       & \begin{tabular}[c]{@{}c@{}}Stochastic\\ Neuron/Synapse\end{tabular}                & \begin{tabular}[c]{@{}c@{}}RRAM Synapse\\ IF neuron\end{tabular}                & \begin{tabular}[c]{@{}c@{}}PCM Synapse\\ LIF neuron\end{tabular} & \begin{tabular}[c]{@{}c@{}}Stochastic\\ Neuron/Synapse\end{tabular}          & \begin{tabular}[c]{@{}c@{}}RRAM Synapse\\ IF neuron\end{tabular} \\ \hline
\textbf{Accuracy}                                                             & \begin{tabular}[c]{@{}c@{}}92.30\%\\ (784 $\times$ 400 $\times$ 10) \\ Trained on 60k \\ MNIST digits\end{tabular} & \begin{tabular}[c]{@{}c@{}}86\%\\ (784 $\times$ 10)\\ Trained on 50k \\ MNIST digits\end{tabular} & N/A                                                              & \begin{tabular}[c]{@{}c@{}}$\approx 88\%$\\(784 $\times$ 500 $\times$ 10) \\Trained on 50k \\ MNIST digits\end{tabular}                                                                       & N/A                                                              \\ \hline
\textbf{Technology}                                                           & 90nm                                                                               & Non-CMOS                                                                        & Non-CMOS                                                         & Non-CMOS                                                                   & 180nm                                                            \\ \hline
\end{tabular}}

\scriptsize

* Measured power: `stochastic bit' + 15b counter + etc. = 226$\mu A$*1.4V*26.7ns = 8.4pJ

** Estimated neuron power: 226$\mu A$ * (33.3/153.2)*1.4V*26.7ns = 1.84pJ

(Post-layout simulated current: 153.2$\mu$ = `stochastic bit'[33.3$\mu$]+ others[119.9$\mu$])

*** Peak power with a single spike duration of $\approx$ 10 $\mu s$

**** Normalized power~\cite{chen2002power} from 180nm to 90nm: 9.3pJ *(90/180)*1.4/1.8=3.61pJ
\end{table}

\subsection{sBSNN for MNIST Digit Recognition}\label{sec:result_system}
The sBSNN implementation was functionally trained and evaluated using the measured neuronal and synaptic dynamics shown in Figs.~\ref{fig:measure_stochastic-bit}(b) and~\ref{fig:measure_pstdp}(b), respectively, on the MNIST digit recognition dataset. The accuracy on the test dataset is 65.88\% for an SNN of 400 excitatory neurons trained on 900 MNIST digit patterns, which was sufficient for all the neurons to learn general input representations as depicted in Fig. \ref{fig:SBSNN_Digits}. Any more increase in the number of training patterns could deteriorate the learnt representations, leading to further loss in accuracy. The accuracy can be improved by increasing the number of excitatory neurons and/or by incorporating an additional fully-connected classification layer trained on a larger fraction of the dataset. We augmented the SNN with a softmax readout layer of 10 neurons corresponding to the 10 classes in the MNIST handwritten digit recognition task, where each readout neuron is fully-connected to all the excitatory neurons. For a given input pattern, the spike count of the excitatory neurons are estimated using the sSTDP trained sBSNN, and subsequently fed to the softmax readout layer, which predicts the test pattern to belong to the category represented by the readout neuron with the highest activation. We trained the readout layer on the entire training dataset using the Adam optimizer \cite{kingma2014adam}, which is a popular gradient-based supervised training algorithm, and cross-entropy loss function with learning rate of 0.001 for 8 epochs. We obtained higher accuracy of 92.30\% on the entire MNIST test dataset of 10,000 images.

sBSNN offers possibility of up to 32$\times$ neuronal and synaptic memory compression relative to similarly sized full precision (32-bit) SNN with accuracy loss that can be minimized for larger SNNs. The energy of the sNeuron with the measurement blocks (refer to the sNeuron measurement setup in Fig. \ref{fig:measure_stochastic-bit}) is measured to be 8.4pJ/spike. The standalone neuronal energy is estimated to be 1.84pJ/spike as detailed in \autoref{tab:comparison}. In addition, \autoref{tab:comparison} also indicates that the proposed implementation offers lower neuronal energy consumption compared to related works in 90nm CMOS process.

\begin{figure}[t]
  \centering
  \includegraphics[width=0.8\columnwidth]{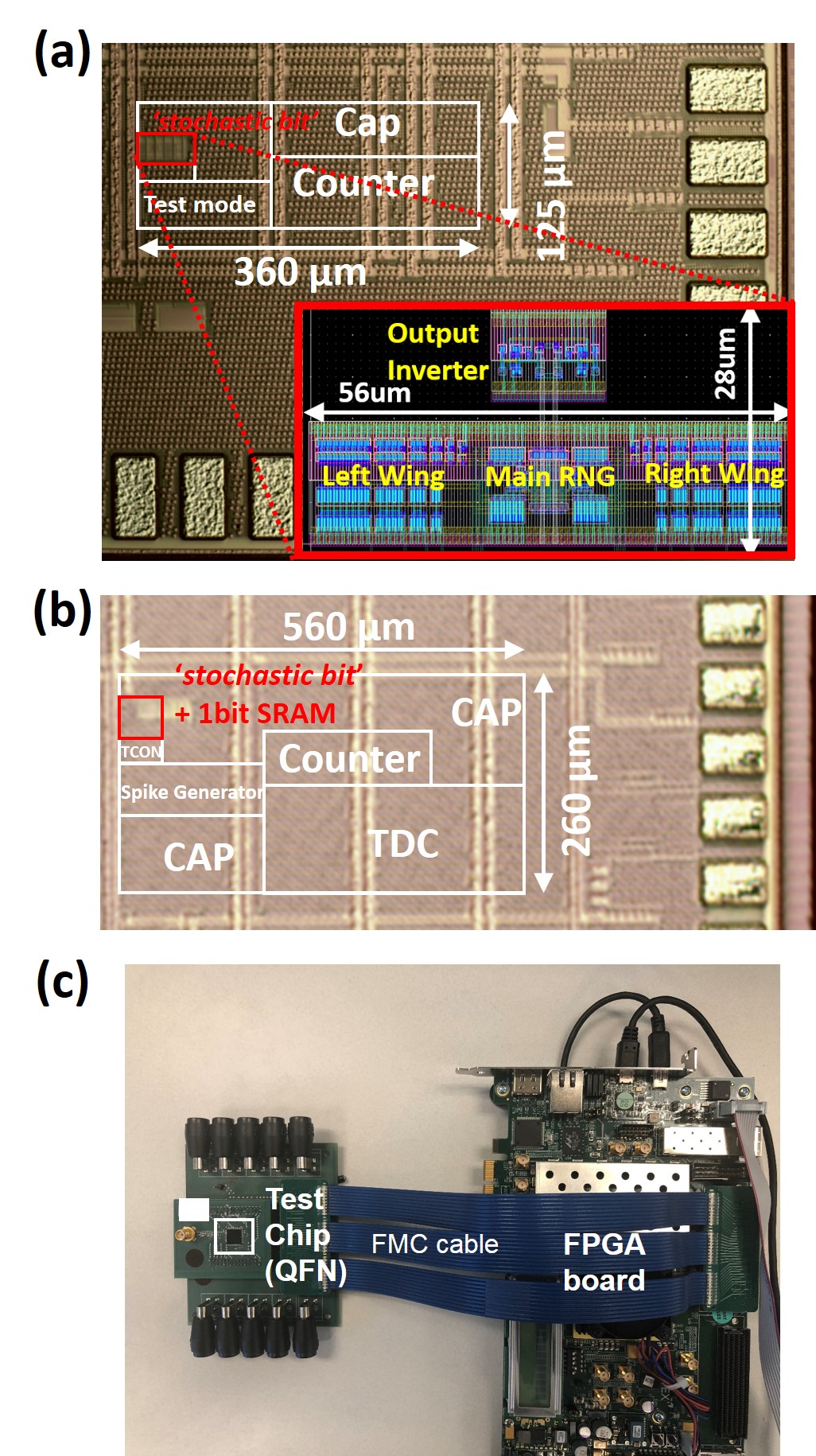}
  %\shrinkBeforeFigureCaption
  \caption{(a) Die photo of the \textit{`stochastic bit'} and its layout (refer to the inset). (b) Die photo of the stochastic binary synapse composed of the \textit{`stochastic bit'} and 6T-SRAM bitcell. (c) Test chip measurement setup using FPGA.}\label{fig:result_measure}
  %\shrinkAfterFigure
\end{figure}

\subsection{Energy efficiency}\label{sec:result_pwr_eff}
Finally, we estimate the energy efficiency of the two-layer sBSNN implementation composed of 784 input and 400 output sNeurons in terms of Tera-operations (TOPS) per Watt. Our functional simulations indicated that the average number of transitions in the AND gate of stochastic synapses is $\sim$700 out of 784$\times$400 total possible transitions. The average power consumed by the AND gate per transition in 90nm CMOS process is estimated as 0.80$\mu W$, which totals to 0.56mW per time step. Every output sNeuron requires a 10-bit ones counter for accumulating the maximum weighted input sum of 784, and the \textit{`stochastic bit'} to spike probabilistically. The average weighted input sum received by the output sNeurons is functionally determined to be 21. The average power consumed by the 10-bit ones counter is estimated to be 0.558mW per sNeuron while that of the \textit{`stochastic bit'} is measured to be 0.033mW per sNeuron. The total output neuronal power is 236mW (0.558mW$+$0.033mW\ $\times$400) while that of the input neurons is 25.87mW (0.033mW$\times$784). The proposed implementation performs 23.52 TOPS (784$\times$400$\times$2$\times$37.5MHz) while consuming 262.8mW, leading to energy efficiency of 89.49TOPS/Watt. The high energy efficiency can be attributed to binary dot product computations and the inherent sparsity in the neuronal spiking activity offered by SNNs. Figs. \ref{fig:result_measure}(a)-(b) show the die shot of the sNeuron, synapse, and the layout of the \textit{`stochastic bit'} core (inset of Fig. \ref{fig:result_measure}(a)). For measurements, we interfaced an FPGA to the QFN packaged chip on a custom PCB as depicted in  Fig.~\ref{fig:result_measure}(c).

\begin{figure}[t]
  \centering
  \includegraphics[width=0.8\columnwidth]{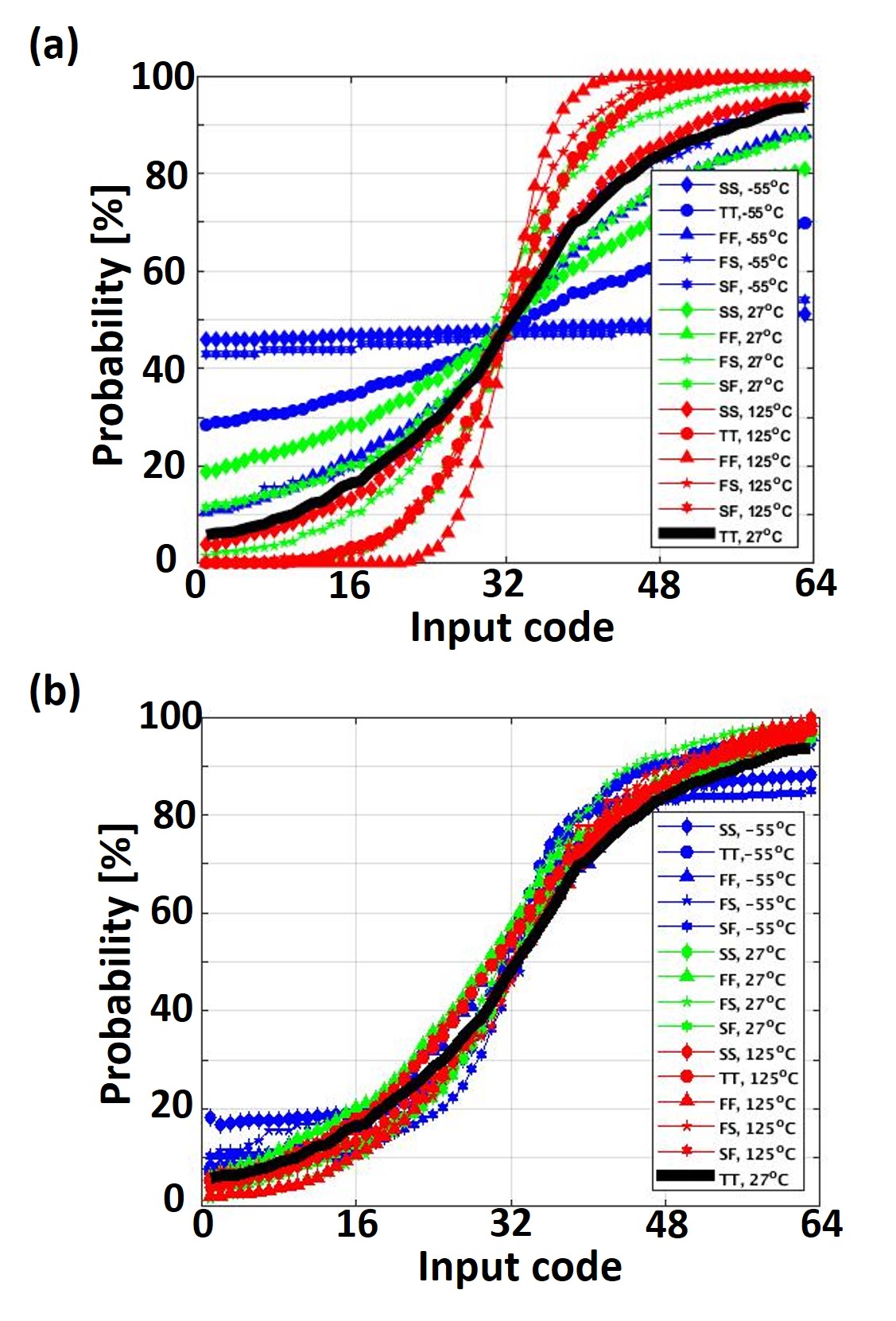}
  %\shrinkBeforeFigureCaption
  \caption{ (a) The switching probability curves with process (FF, TT, SS, FS, SF) and temperatures (-55 $^{\circ}C$, 27 $^{\circ}C$, 125 $^{\circ}C$) variations. (b) The compensated switching probability curves for all corners presented in (a).}\label{fig:result_variation}
  %\shrinkAfterFigure
\end{figure}

%\revised{
\subsection{Process and temperature variation}\label{sec:result_variation}
Fig. \ref{fig:result_variation}(a) shows the simulated switching probability curves affected by process and temperature variations. The black solid line represents the baseline of our design and the other lines represent variations caused by the different combinations of process corners (FF, TT, SS, FS, SF) and temperatures (-55$^{\circ}C$, 27$^{\circ}C$, 125$^{\circ}C$). 
The (SS, -55$^{\circ}C$) corner shows less than 10\% change in probability due to decreased temperature and current, decreasing noise or the source of the randomness. 
The variations can be easily compensated by having variable size of M1 and M2 transistors of Fig. \ref{fig:stochastic-bit}(b) in the same way we size M3 and M4 transistors. The size ratio between M1/M2 transistors and M$_{l1x}$/M$_{r1x}$ transistors determines the unit step change of probability and thus, the slope of the probability curve. 
Fig. \ref{fig:result_variation}(b) shows the compensated switching probability curves from all corners presented in Fig. \ref{fig:result_variation}(a). 
In addition to the variation compensation, this approach also allows us to control the shape and slope of the probability curve at the cost of area required for sizing M1 and M2 transistors. 
Further, the probability range can also be controlled through the NMOS codes applied to M3 and M4 transistors as shown in Fig. \ref{fig:measure_stochastic-bit}(c). 
%}

%% file: sec/5-conclusion.tex
\section{Conclusion}\label{sec:conclusion}

We proposed \textit{`stochastic bit'} enabled Binary SNN (sBSNN), composed of stochastic spiking neurons (sNeurons) and binary synaptic weights, for energy- and memory-efficient neuromorphic computing at the edge. sBSNN computes probabilistically with only one-bit precision for both the constituting sNeurons and synapses, leading to on-chip visual image recognition with compressed memory requirement. We presented an energy-efficient implementation of two-layer fully-connected sBSNN using \textit{`stochastic bit'} as the core computational primitive to realize the sNeurons and binary synapses (stochastic during training and deterministic during inference) fabricated in 90nm CMOS process. We demonstrated memory-efficient on-chip training of the binary synaptic weights using the stochastic-STDP (sSTDP) training algorithm. The proposed implementation, by virtue of sparse event-driven computing enabled by state-less sNeurons and binary weights, offered high energy-efficiency of 89.49 TOPS/Watt. We believe that sBSNN can provide a promising solution for building the next generation of intelligent devices capable of real-time on-chip learning.

%% file: sec/6-acknowledgement.tex
\section{Acknowledgment}\label{sec:acknowledgement}

This work was supported in part by the Center for Brain Inspired Computing (C-BRIC), one of the six centers in JUMP, a Semiconductor Research Corporation (SRC) program sponsored by DARPA, by the Semiconductor Research Corporation, the National Science Foundation, Intel Corporation, and the DoD Vannevar Bush Fellowship.